# Self-Stabilizing Task Allocation In Spite of Noise

Anna Dornhaus[1], Nancy Lynch[2], Frederik Mallmann-Trenn[*,2], Dominik Pajak[2], and Tsvetomira Radeva[2]

[1]University of Arizona, Department of Ecology and Evolutionary Biology
[2]MIT, CSAIL

## Abstract

We study the problem of distributed task allocation inspired by the behavior of social insects, which perform task allocation in a setting of limited capabilities and noisy environment feedback. We assume that each task has a *demand* that should be satisfied but not exceeded, i.e., there is an optimal number of ants that should be working on this task at a given time. The goal is to assign a near-optimal number of workers to each task in a distributed manner and without explicit access to the values of the demands nor the number of ants working on the task.

We seek to answer the question of how the quality of task allocation depends on the accuracy of assessing whether too many (`overload`) or not enough (`lack`) ants are currently working on a given task. Concretely, we address the open question of solving task allocation in the model where each ant receives feedback that depends on the *deficit* defined as the difference between the optimal demand and the current number of workers in the task. The feedback is modeled as a random variable that takes values `lack` or `overload` with probability given by a sigmoid of the deficit. Each ants receives the feedback independently. The higher the overload or lack of workers for a task, the more likely is that an ant receives the correct feedback from this task; the closer the deficit is to zero, the less reliable the feedback becomes. We measure the performance of task allocation algorithms using the notion of *regret*, defined as the absolute value of the deficit summed over all tasks and summed over time.

We propose a simple, constant-memory, self-stabilizing, distributed algorithm that converges from any initial distribution to a near-optimal assignment. We then show that our algorithm works not only under stochastic noise but also in an adversarial noise setting. Finally we prove a lower bound on the regret for any constant-memory algorithm, which matches, up to a constant factor, the regret achieved by our algorithm.

---

[*]mallmann@mit.edu

# Contents

# 1 Introduction

Task allocation in social insect colonies is the process of assigning workers to tasks such as foraging, scouting, nursing, *etc.* in a way that maximizes the reproductive success of the colony. Each ant can work on each of the tasks but the demands of the tasks are different and might also vary over time. The ants probably neither know the number of ants needed nor can count the current number of ants working on a given task [20]. Therefore the allocation has to be performed based only on the feedback from the tasks. The feedback, in biology called 'task stimulus', corresponds for example to sensing a too-high temperature in the nest, seeing light through a hole in the nest-wall, or smelling a pheromone produced by hungry brood. Despite using limited communication, local observations, and noisy sensing, many ant species are known to excel at task allocation. How do ants perform task allocation and what can we learn from their behavior?

In [11], the authors proposed a solution to the problem of task allocation in the case where the feedback received by the ants is always correct. More precisely, in [11], each task $j \in [k]$ has a demand $d^{(j)}$, that represents the number of workers needed at that task. If the *load, i.e.,* the number of ants working on the task, exceeds the demand, then all ants receive feedback `overload`. Conversely, if the load is below or equals the demand of the task, then all ants receive feedback `lack`. Such a feedback function is rather unrealistic in ant colonies due to its sharp transition between `overload` and `lack`— it requires each ant to be able to tell the difference between $d^{(j)}$ and $d^{(j)} + 1$ number of workers at a task. The authors in [11] therefore pose the open problem of considering a *weaker*, noisy version of the binary-feedback—the focal point of this paper.

We study the performance of task allocation in two noise models: a realistic, stochastic model and a more theoretical, adversarial model. In the stochastic model, the feedback from the tasks for each ant in each step is a random variable with possible values `lack` and `overload`. The probability that it takes value `lack` equals to the sigmoid function of the deficit (demand minus load[1]) of the task. In the adversarial noise model, the feedback is deterministic and it is always correct if the absolute value of the deficit is large but can be arbitrary if the absolute value of the deficit is small.

We assume that the ants regularly receive the feedback from all the tasks. However, there is a delay between the moments when an ant collects the feedback and when it changes its allocation during which other ants may also make some decisions. To model this delay we assume, similarly to [11], synchronous rounds: at the beginning of each round each ant receives binary feedback of the load of the task. The ants then concurrently make a decision of whether to join or to leave their current task.

How can the ants make independent decisions and achieve a 'good' task allocation in spite of *outdated observations*, in spite of *noise* and in spite of the *lack of global information*—not knowing the demands nor the current load of a task?

In order to define what a 'good' task allocation means we propose to use the notion of *regret*. Intuitively, the regret in our setting measures the sum over all rounds over all tasks of the absolute value of the deficits (demand minus load) of the tasks. Here we penalize overload and underload equally: An underload corresponds to work that is not being done, and each ant exceeding the demand of a task corresponds to work being wasted (or, even worse, sometimes the excessive number of workers in a task may block each other and decrease the efficiency [15, 12]). Note, that we do not charge any cost for switching tasks.

It turns out that any constant-memory algorithm for task allocation can be lower bounded in terms of this 'cost' function. We show that the quality of task allocation is determined by the *critical value* (see Section 2 for a formal definition). Intuitively, the critical value determines a value of the deficit (seen as a fraction of the demand) for which the feedback is correct for each ant with high probability. This corresponds to a smallest value of the deficit at which the sigmoid is very close to 1 and a largest value for which it is very close to 0. The lower bound shows that the regret of the optimal constant-memory algorithm is linear in the sum of demands times this critical value.

---

[1]Note, that the deficit can be negative.

We then provide a simple, constant-memory Algorithm Ant that utilizes the oscillations in the number of workers in each task in order to achieve a stable allocation in the synchronous model. The size of the oscillation at each task in our algorithm is proportional to the demand of the task times the critical value. In the light of our lower bound the algorithm achievess a constant factor approximation of the optimal regret. Our algorithm assumes a slightly stronger form of synchronization where the ants take actions through cycles (phases) of length two (for example, day and night) and all of the ants are at the same step of the cycle[2]. Our algorithm is parameterized by a *learning parameter* $\gamma$ that upper bounds the critical value. Smaller values of $\gamma$ yield better bounds for the regret but slower convergence time.

On a high level, in our Algorithm Ant, the ants repeatedly collect two samples each to asses which decision (joining, leaving or remaining) would be optimal in terms of regret. To obtain the first sample, all ants assigned to a task work on it. For the second sample ants collectively leave their tasks with a small probability. If both samples of a working ant indicate an overload, then it will permanently leave the task with a small probability. If both samples of an idle ant indicate underload, then such ant will join this task (or if there are multiple such tasks then the idle ant joins one chosen uniformly at random). This two-sample approach can be seen as computing the slope of the regret function in each task. The ants collectively, in a distributed fashion, increase or decrease the loads in the tasks by a small fraction depending on the computed "direction of the gradient" and learning parameter $\gamma$. The algorithm bears similarities to gradient descent; in our case a noisy and distributed version. We show that this algorithm achieves a task allocation that is up to a constant factor optimal with respect to the aforementioned lower bound.

We then build upon our two-sample algorithm to derive a more theoretical algorithm Algorithm Precise Sigmoid that uses more memory as well as longer synchronous phases (consisting of more than two rounds). This algorithm, together with a lower bound, establish a theoretical tradeoff between the memory available at each ant and the optimal regret that can be achieved by an algorithm using this memory. Moreover, we also consider the adversarial noise setting in which we propose algorithm Algorithm Precise Adversarial that also achieves a constant approximation of the best possible regret.

We also show that if the deficit is too close to 0 for some number of steps then (due to the noisy feedback) it drastically increases. It is therefore impossible to keep the absolute value of the deficit very small for too long. This means that small oscillations in the number of workers at each task are unavoidable in any algorithm that 'tries' to achieve a good and stable allocation. Our proposed solution is to avoid getting too close to 0 with the absolute value of the deficit and use the oscillations (jumping between positive and negative deficit) to achieve stable allocation and asymptotically optimal regret.

## 1.1 Related work

Distributed task allocation in the context of social insect colonies and similar simple models have been studied for years in both the theoretical distributed computing and the behavioral ecology communities.

The most related previous work on task allocation is [11], in which the authors also assume synchronous rounds and binary feedback. The authors present a very simple algorithm that converges to an almost-optimum allocation (the allocation that differs from the demand by at most 1 at each task) and analyze its convergence time. Considering a noisy version of the model was left as an open question.

Moreover, the author of [29] provide a model similar to that of [11] but they also study different versions of the feedback that ants receive from the environment, which varies in the amount of information the ants receive about the deficits of the tasks. In short, the model consists of two feedback components: a *success* component that informs each ant in each round whether it is successful, e.g., needed at the task it is currently working on, and a *choice* component that provides unsuccessful ants with an alternative task to work on. The results in [29] analyze the convergence time of task allocation, and as such are not directly comparable to our work here. In [29], the noise model is very rudimentary: in each round the feedback of the binary success component can be noisy for at most a small fixed number of ants. The results do not generalize to our setting.

---

[2]In a theoretical model, this can be achieved using one bit of additional memory for each worker and using very limited communication, *e.g.* [4].

The problem of task allocation in social insect colonies has been well studied in the communities of theoretical and experimental biology. The observations show that social insect colonies are self-organized, with no individuals directing the task choices of others, with interactions between individuals potentially affecting task selection [18]. Workers in a colony may switch tasks as needed [17], although this may come at additional cost [25]. The concept of task switching gives rise to an intriguing question: what is the algorithm used by the ants to decide whether to switch and which tasks to choose? Some notable examples of models of task allocation [3] include (1) the threshold-based model where ants compared the stimulus of a task to their built-in threshold to determine whether to work on a given task, and (2) the 'foraging for work' model [33] where the ants are believed to actively look for work when they are idle or redundant in the current task. In some species the ants are believed to choose the tasks based on physical suitability (*physical polyethism*) [24], whereas in other species, the ants are physically similar and suitable to do any task (*temporal polyethism*) [5]. In this paper, we assume that the ants are identical (no thresholds) and universal (they can work on any task).

Some biological studies have focused on the efficiency of the task allocation process itself, and how it is determined by the specific algorithm used by the ants. For example, [27] and [13] model task allocation determined by social interactions and response thresholds, respectively, and both demonstrate that perfect task allocation of workers to tasks cannot be achieved, potentially due to the speed and accuracy of task allocation trading off against each other. In [32], an algorithm of task allocation is analyzed in a setting where there are no thresholds, the tasks are arranged in a line and there is no noise in sensing of the demand. The goal of [32] is to explain the experimental observations where certain tasks were preferred by older ants.

Additional factors such as individual experience, interactions with other workers, spatial and hierarchical position in the colony, and random encounters with tasks are also known to affect the specific task allocation mechanism employed [6, 14, 18]. Unfortunately, most often it is not precisely known what is the actual algorithm that the ants use to select tasks or how the factors listed above interact to produce variation in preferences across tasks or across individuals [28].

A key property that we observe in our results—oscillations in the task allocation behavior of ants—is also a commonly observed biological phenomenon more generally known as cyclical activity patterns [9]. Although the role of cyclical activity patterns is not completely understood [10], several studies make conjectures that may be related to the conclusions in our paper. First, our assumption that ants perform actions in synchronized rounds and phases as a means of introducing 'delay' between one another's actions is also observed in biological studies. Ants perform actions in bursts of activity and inactivity in order to clear stale information from spreading through the colony [30]. Second, our results suggest that, assuming that ants have constant memory, and noisy environmental feedback, the oscillations are inevitable as the deficit becomes small. We conjecture that such cyclic activity patterns (switching between different tasks and being idle) are necessary and a product of the limitations of the ants and the noisy feedback about number of workers at a task.

Another key assumption we make is that the noise follows a sigmoid function (also known as a *logistic sigmoid activation function*). Such functions appear in countless biological contexts (*e.g.* [16, 22, 31]), to model the uncertainty with which the ants sense the need for work at different tasks. We believe that the versatility and applicability to the real-world problems of the sigmoid noise model makes it a good choice to model the noise of the environment in our setting.

Finally, somewhat related load-balancing processes have been studied under the term *user-based migration* in which the tasks move in a network of resources [1, 23, 2] by querying the load of the current resource and moving to a neighbor in case of an overload. However, this line of research assumes that each resource knows an upper bound on how many tasks it can accept. Furthermore, the setting is noise-free and we hope that this paper can be used to derive more realistic models, in which the load cannot be determined precisely as it is unclear how long each task will need to be processed.

## 2 Model

The sections consists of three parts. First, we present the model, the assumptions and the bulk of the notation. Second, we define the noise models (adversarial noise model and sigmoid noise model). Finally, we define the regret metric and the closeness of a task allocation.

### 2.1 Notation and Assumptions

We have a collection of $n$ ants and $k$ tasks where each task $j \in [k]$ has a fixed demand $d^{(j)}$ meaning that the task requires $d^{(j)}$ many ants assigned to it. Let $\mathbf{d}$ be the demand vector. Let $W_t^{(j)}, j \in [k]$ denote the load of resource $j$ at time $t$, *i.e.,* the number of ants performing the task. For task $j \in [k]$ we define the deficit as $\Delta_t^{(j)} = d^{(j)} - W_t^{(j)}$ and a negative deficit signifies an overload. Unless specified otherwise, each ant has memory that is linear in the number of tasks, but independent of $n$. We assume there is no communication among the ants.

We assume synchronous rounds each consisting of two sub-rounds. We define round $t \geq 1$ to be the time interval between times $t-1$ and $t$. Fix round $t$. In sub-round 1 (of this round), each ant receives noisy feedback of the load situation at time $t-1$ (see definition below). Then, in sub-round 2—based on the obtained feedback and the internal state (memory) of the ant—the ant decides whether to work and on which task during round $t$. We will bundle two consecutive rounds into a phase and we assume that each phase starts for all ants at the same time step.

We assume that the demands do not change, but our results trivially extend to changing demands due to the self-stabilizing nature of our algorithms.

We now give our assumptions on the demand vector. First, we require the demands to be at least of logarithmic size in the number of ants. Second we assume that there is is sufficient slack of the demands, so that not all ants are required to work. Indeed, it is commonly observed that a large fractions of the ants do not work (*e.g.* [7, 8]).

**Assumptions 2.1.** *Assume that for all $j \in [k]$ we have $d^{(j)} = \Omega(\log n)$ for $j \in [k]$. Moreover, assume that the sum of the demands satisfies $\sum_{j \in [k]} d_j \leq n/2$.*

We model ants as finite state automata and assume that the states of all the ants must be always reachable from each other. We do not allow for example algorithms where an ant working on some task can never leave this task. It is well-known (*e.g.* [19]) that ants do not stabilize to a fixed allocation but switch between the tasks whenever it is needed.

**Assumptions 2.2.** *We assume that for any pair of states $s_1, s_2$ (*idle *or working on one of the tasks $j \in [k]$), there must exists a finite sequence of feedbacks $\mathbf{F}_1, \mathbf{F}_2, \ldots$ such that an ant in state $s_1$ transitions with nonzero probability to state $s_2$.*

We use the shorthand w.p. for 'with probability'. We say an event happens w.h.p. 'with high probability' to mean that the event happens w.p. at least $1 - O(1/n)$. We say an event happens with *overwhelming probability* is it happens w.p. at least $1 - e^{-\Omega(n)}$. We define the *configuration* at time $\tau$ to be the internal state of all ants at time $\tau$.

## 2.2 Noisy feedback

We seek to model the noise in the sensing such that the following axioms are fulfilled. First, in case of a very large deficit almost all ants will notice this (w.h.p. all ants receive feedback `lack`). Similarly, in case of a high overload (negative deficit) almost all ants will notice this as well. Second, whenever exactly the correct number of ants are working on a given task, then the 'uncertainty' in this task is the largest and each ant receives `lack` and `overload` with equal probability. In the following we define the sigmoid feedback and the adversarial feedback that both fulfill these axioms.

**Sigmoid feedback** The noisy feedback is modeled by a sigmoid function
$$s(x) = \frac{1}{1+e^{-\lambda x}},$$
for fixed $\lambda \in \mathbb{R}$. At the beginning of round $t$, each ant $i$ receives for each task $j$ noisy feedback $F_t^{(j)}(i)$ of the deficit:

$$F_t^{(j)}(i) = \begin{cases} \texttt{lack} & \text{with probability } s(\Delta_{t-1}^{(j)}) \\ \texttt{overload} & \text{otherwise} \end{cases}.$$

We write $\mathbf{F}_t(i)$ to denote the vector of feedbacks for all tasks that ant $i$ receives at time $t$. It is not crucial for our results to have a sigmoid function; in fact all our results apply for any monotone antisymmetric function $s$ with exponential decay and $\lim_{x \to -\infty} s(x) = 0$ and $\lim_{x \to \infty} s(x) = 1$ and $s(0) = 1/2$.

**Adversarial feedback** This feedback is parameterized by threshold $\gamma^{ad} \in \mathbb{R}$. The adversarial feedback $F_t^{(j)}(i)$ received at the beginning of round $t$ by each ant $i$ for each task $j$ is defined as:

$$F_t^{(j)}(i) = \begin{cases} \texttt{lack} & \Delta_{t-1}^{(j)} > \gamma^{ad} d^{(j)} \\ \text{arbitrary value in } \{\texttt{lack}, \texttt{overload}\} & \Delta_{t-1}^{(j)} \in [-\gamma^{ad} d^{(j)}, \gamma^{ad} d^{(j)}] \\ \texttt{overload} & \Delta_{t-1}^{(j)} < -\gamma^{ad} d^{(j)} \end{cases}.$$

**Critical value** The critical value is the deficit which ensures that all the ants receive with high probability the correct feedback provided that the overload/lack is far enough away from 0. The idea is that an ant will be able to see an overload (lack, respectively) if the number of ants working at a task exceeds $(1+c)d$ (or is lower than $(1-c)d$, respectively) for some small constant $c > 0$ and demand $d$. The value $c$ is the critical value. Note that, by symmetry, a deficit of $-cd$ also ensures that all ants receive feedback `overload`.

**Definition 2.3** (critical value and grey zone). *We define the* critical (feedback) value *for the respective models as follows.*

- *Sigmoid feedback model. Let $y(x) = \min_{x' \in \mathbb{R}} \{s(-x' \cdot d^{(j)}) \leq x \colon \text{ for all } j \in [k]\}$. We define the critical (feedback) value to be $\gamma^* = y(1/n^8)$. Observe that, due to the antisymmetry of the sigmoid, $s(-\gamma^* \cdot d^{(j)}) = 1 - s(\gamma^* \cdot d^{(j)})$. Wwe make no assumption on $\gamma^*$ apart from being smaller than $1/2$.*

- *Adversarial feedback model. We define the critical (feedback) value $\gamma^*$ to be $\gamma^{ad}$.*

*We define for each task $j \in [k]$ the* grey zone *to be $g_j = [-\gamma^* d^{(j)}, \gamma^* d^{(j)}]$.*

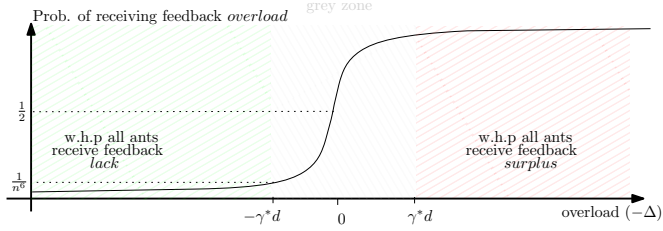

Figure 1: Whenever the overload is in the green (red, respectively) region, all ants will receive w.h.p. the feedback `lack` (`overload`, respectively). Whenever the overload is in the grey region (falling tilling pattern), and the closer the overload is to 0, the more unpredictable is the feedback received by the ants.

### 2.3 Regret metric, closeness and farness.

In order to compare the quality of task allocation strategies we use the following notion of regret $r(t)$, defined for time $t \in \mathbb{R}$ as:
$$r(t) = \sum_{j \in [k]} |d^{(j)} - W_t^{(j)}| = \sum_{j \in [k]} |\Delta_t^{(j)}|.$$
Furthermore, we define the (total) regret up to time $t$ as:

$R(t) = \sum_{i \leq t} r(i)$.

The idea behind the regret metric is that if over large scales of time the deficit is only rarely large, then this should not affect the performance (*i.e.*, the survival of the colony) too much. Moreover, as we will see later, any constant-memory algorithm will have oscillations (or a significant deficit) and hence classical metrics such as convergence time carry little meaning. It is worth pointing out that we penalize lack and overload (measuring a waste of energy) equally and we leave it as a future direction to use different weights.

We say that an algorithm produces an assignment that is *c-close* if the average regret of the allocation is at most $c$ times the critical value and the sum of demands, *i.e.*, $\lim_{t \to \infty} \frac{R(t)}{t} \leq \left( c\gamma^* \sum_{j \in [k]} d^{(j)} \right) + O(1)$. Similarly, we say an the assignment of an algorithm is *c-far* if the average regret of is at least $c$ times the critical value and the sum of demands, *i.e.*, $\lim_{t \to \infty} \frac{R(t)}{t} \geq \left( c\gamma^* \sum_{j \in [k]} d^{(j)} \right)$.

## 3 Results and Discussion

In this section we present our results. We start with an overview of all the results. Second, we give results that hold for both noise models (sigmoid and adversarial). We then give the results for the sigmoid noise model followed by the results for the adversarial noise model. The corresponding algorithms are defined in Section 4, Section 5 and Appendix C.

### 3.1 High-Level Overview

Our results on sigmoid noise are threefold.

(i) In our main theorem we show that the somewhat realistic Algorithm Ant with parameter $\gamma$ results w.h.p. in an assignment that is $5\frac{\gamma}{\gamma^*}$-close provided that $\gamma \geq \gamma^*$. In particular, the assignment is 5-close if $\gamma = \gamma^*$. (Theorem 3.1)

(ii) Second, we develop Algorithm Precise Sigmoid that is slightly more artificial but achieves arbitrary closeness: for any $\varepsilon \in (0, 1)$, the algorithm produces w.h.p. an assignment that is $\varepsilon$-close provided that $\gamma = \gamma^*$. The algorithm uses $O(\log(1/\varepsilon))$ memory and synchronous phases of length $O(1/\varepsilon)$. (Theorem 3.2)

(iii) Finally, we show that this bound is in fact tight: any algorithm using $c\log(1/\varepsilon)$ memory, for small enough constant $c$ will produce, with overwhelming probability, an assignment that is at least $\varepsilon$-far for almost all time steps. (Theorem 3.3)

In the adversarial noise setting, we prove the following bounds.

(i) Our Algorithm Ant also achieves in the adversarial noise setting w.h.p. an assignment that is $5\frac{\gamma}{\gamma^*}$-close. (Theorem 3.1)

(ii) Second, we develop Algorithm Precise Adversarial that is slightly more artificial but achieves arbitrary closeness: for any $\varepsilon \in (0,1)$, the algorithm produces w.h.p. an assignment that is $(1+\varepsilon)$-close provided that $\gamma = \gamma^*$. The algorithm uses $O(\log(1/\varepsilon))$ memory and synchronous phases of length $O(1/\varepsilon)$. (Theorem 3.6)

(iii) Any algorithm, regardless of the memory, produces an assignment that is in expectation at least $(1 - o(1))$-far. (Theorem 3.5)

## 3.2 Results for Both Noise Models

We proceed by giving the precise statements of Theorem 3.1 showing that the assignment of Algorithm Ant is w.h.p. $5\frac{\gamma}{\gamma^*}$-close.

**Theorem 3.1.** *The following holds for the sigmoid noise model as well as for the adversarial noise model. Consider an arbitrary initial allocation at time $0$. Fix an arbitrary $t \in \mathbb{N}$. Algorithm Ant with learning rate $\gamma \geq \gamma^*$ has w.h.p. a total regret during the first $t$ rounds that is bounded by*

$$R(t) \leq c\frac{nk}{\gamma} + \left(5\gamma \sum_{j \in [k]} d^{(j)} + 3\right) \cdot t,$$

*for some constant $c$.*

*Moreover for any interval of time of length at most $n^4$, for each task $j \in [k]$, the absolute value of the deficit is w.h.p. in all but $O(k \log n/\gamma)$ rounds bounded by $5\gamma d^{(j)} + 3$.*

Note that we allow $t$ to take arbitrary values—in particular, values that are super-exponential in $n$.

## 3.3 Results for Sigmoid Noise

In the following we show that Algorithm Precise Sigmoid is $\varepsilon$-close but requires synchronous cycles of length $O(1/\varepsilon)$. For applications where this is tolerable, the algorithm becomes superior and in fact optimal w.r.t. to the closeness as we will see later.

**Theorem 3.2.** *The following holds for the sigmoid noise model. Fix an arbitrary $\varepsilon = \Omega(1/\log n)$. Algorithm Precise Sigmoid with parameter $\gamma \geq \gamma^*$ and $\varepsilon$ has a regret of*

$$\lim_{t \to \infty} \frac{R(t)}{t} = \gamma\varepsilon \sum_{j \in [k]} d^{(j)} + O(1),$$

*where we suppress all terms that are independent of $t$. The algorithm uses $O(\log(1/\varepsilon))$ memory and synchronous rounds of length $O(1/\varepsilon)$.*

We finally show a lower bound on the memory size for any algorithm that is $\varepsilon$-close establishing optimality of Algorithm Precise Sigmoid.

**Theorem 3.3.** *The following holds for the sigmoid noise model. Assume $\gamma^* \leq 1$ and let $c \leq 1$ be a small enough constant. Let $\varepsilon \in (0, 1/4)$ and let $n$ (number of ants) be large enough integer. There exists a demand vector $(d^{(1)}, d^{(2)}, \ldots, d^{(k)})$ such that for any collection of $n$ ants executing (possibly distinct) algorithms $A_1, A_2, \ldots, A_n$, each using at most $c\lfloor\log(1/\varepsilon)\rfloor$ bits of memory, the following holds. For any time $t \geq 1/\sqrt{\varepsilon}$ with overwhelming probability*

$$R(t) \geq \left(\varepsilon \gamma^* \sum_{j \in [k]} d^{(j)}\right) t.$$

*Furthermore, if the deficit for all tasks is below $2\varepsilon\gamma^* d^{(j)}$ for a constant number of consecutive steps (of sufficient length), then, with overwhelming probability, there will be a task $j$ with an oscillation (i.e., a deficit) of order $\omega(\gamma^* d^{(j)})$.*

In the proof of our lower we show that, unless the deficit is of order $\gamma^* \sum_{j \in [k]} d^{(j)}$ oscillations are unavoidable.

The lower bound focuses on the perpetual cost of the algorithm that, once $t$ is large enough, marginalizes the initial costs. Nevertheless, we point out that if the demands can change arbitrarily in any round, then any algorithm that does not know when the demands, will pay an initial regret of $\Omega(n)$.

**Remark 3.4.** *The guarantees from Theorem 3.1 and Theorem 3.2 even apply if the feedback is arbitrarily correlated as long as the marginal probability for each ant to receive incorrect feedback outside the grey zone is $1/n^c$ for some small constant $c > 1$.*

*Moreover, our algorithm trivially also works—due to its self-stabilizing nature—for changing demands.*

*Furthermore, even though we state Algorithm Ant and Algorithm Precise Sigmoid in a way that the feedback of all the tasks is collected (as in [11]), this is not necessary and only the initial cost would change if each ant could only receive feedback from one (adaptively) chosen task. In addition for the ease of presentation, our algorithms store the samples inefficiently (e.g. the median). The bounds we claim on the space complexity assume slightly smarter, but obvious techniques.*

*Finally, the required bound on the sum of demands of Assumptions 2.1 can be relaxed and the constant need not be $1/2$; it must only guarantee that $\sum_{j \in [k]}(1 + 5\gamma^*)d^{(j)} \leq c^*n$ for some constant $c^* < 1$.*

### 3.4 Results for Adversarial Noise

In the following we present our results for the adversarial noise model.

**Theorem 3.5.** *In the adversarial noise model, any algorithm, possibly randomized, using unlimited memory and communication has expected regret of at least:*

$$\mathbb{E}[R(t)] \geq (1 - o(1))t\gamma^* \sum_{j \in [k]} d^{(j)}.$$

Note that in the sigmoid noise model a $\varepsilon$-closeness is possible showing a separation between the two models. Using a modification of Algorithm Ant we can achieve $(1 + \varepsilon)$-closeness.

**Theorem 3.6.** *Let $\gamma^* \leq 1/4$. Fix an arbitrary $\varepsilon = \Omega(1/\log n)$. Algorithm Precise Adversarial with parameter $\gamma \geq \gamma^*$ using synchronized cycles of length $O(1/\varepsilon)$ and memory $O(\log(1/\varepsilon))$ achieves in the adversarial noise model the following asymptotic bound on regret:*

$$\lim_{t \to \infty} \frac{R(t)}{t} = \gamma(1 + \varepsilon) \sum_{j \in [k]} d^{(j)} + O(1).$$

Interestingly, our algorithm also minimizes the total number of switches of ants between tasks in comparison to Algorithm Ant; this might be of interest, if one changes the regret to incorporate costs for switching between tasks.

# 4  Algorithm Ant - Theorem 3.1

In this section we introduce Algorithm Ant. We start by giving the intuition and we refer the reader to page 11 for full definition of the algorithm. We present the intuition behind the algorithm by focusing on the single task setting, however, the full algorithm is written and analyzed in the general setting with $k > 1$. We divide time into phases, where each phase consists of two consecutive rounds. This allows each ant in each phase to take two *samples* where a sample is simply the binary feedback (lack or overload) for the task. We say that the samples are taken at different points which means that the number of ants working at the considered task is temporarily reduced between the first and the second sample. The intuition of taking two samples is as follows. If both samples indicate an overload, then since the second sample is taken while the number of workers is reduced (and it still shows overload), we should decrease the number of workers in the task. If one sample indicates overload and the other lack, then load is likely to be close to the optimal demand and there is no need to change the number of ants working on the task in the next phase. Finally, if both samples indicate a lack, then additional ants are required to join the task.

How can we, without any central control, take two samples of feedback for different values of load? The ants achieve it by independently with probability $\Theta(\gamma)$, temporarily, pausing their work on the current task. With this idea we can obtain two samples that are taken at two different values of load in the task, and the distance between the samples can be regulated by the probability of pausing. More precisely, assume that $\mathcal{W}$ ants are working on some fixed task with demand $d$ in the first round of some phase. Then the first sample is simply the feedback of this load (*i.e.*, lack with probability $s(\mathcal{W} - d)$ and overload otherwise). The second sample (in the second round of the phase) is taken after each ant independently stopped working with probability $c_s \gamma$, where $c_s$ is a constant. Therefore, the number of ants working in the second sample is roughly $\mathcal{W}(1 - c_s \gamma)$ (see Figure 2 for an illustration).

Why do we need to take two samples that are 'spaced' apart? The reason is that if the deficit of task $j \in [k]$ in the current step is in the range $[-\gamma^* d^{(j)}, \gamma^* d^{(j)}]$ ("grey zone") then the feedback in this step might be unreliable. In our algorithm we take two samples—spaced far enough apart—so that at least one of them must lie w.h.p. outside of the above range. When the deficit is outside this range, then with high probability, all the ants receive the same, correct feedback. If $\mathcal{W} \geq d(1 + \gamma^*)$ then the first sample shows overload hence after this phase we may decrease the number of ants in this task but we cannot increase (to increase we need lack in both samples). Similarly if $\mathcal{W} \leq d(1 + \gamma^*)$ then the second samples shows lack (because the samples are sufficiently spaced apart) hence we cannot decrease but we may increase in this phase. This shows that, informally speaking, the load of a task w.h.p. can move only in the correct direction (if we look only at the loads after the decision at the end of each phase and disregard the loads in the second sample in each phase). Finally we can identify a "stable zone" in which the first sample lies on the right side of grey zone (Figure 2) and the second sample lies to left of the grey zone. Note that once the load of a task is in the stable zone, it can neither increase nor decrease (because the first sample always gives overload and the second lack).

**Analysis**  The idea of our proof is to analyze the regret in three different regions. For this we split the regret $R(t)$ into $R^+(t), R^{\approx}(t)$ and $R^-(t)$ which intuitively measure the regret in the overload region, the region that is close to 0 deficit and in a region in which there is a lack of ants.

We divide time into intervals of length $n^4$ and show, by using a series of potential arguments, that $R^-(t)$ is w.h.p. bounded by $O(nk)$ during any fixed interval. We do this by arguing that in any phase w.h.p. either the 'lack' of the resources decreases by a significant amount or the number of tasks that are underloaded does. It turns out that both quantities are monotonically decreasing every second round and will w.h.p. quickly reach 0. We then bound $R^+(t)$, by showing, that w.h.p. each task can only cause at most once in the interval an overload of $\Theta(n)$ and once this happens, the overload of the task will decrease in a geometrical fashion. Summing over all tasks gives w.h.p. a total regret of

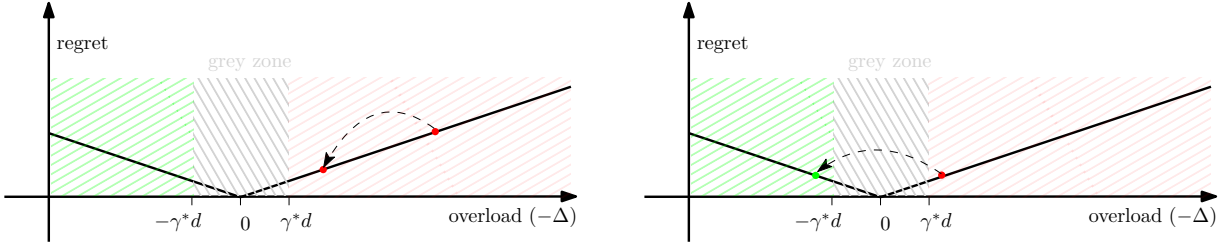

Figure 2: A typical execution of one phase in which two samples are taken—each ant temporarily drops out with probability linear in $\gamma \approx \gamma^*$. It is convenient to think of the process as gradient descent-like with learning rate $\gamma$. The left figure shows a phase in which both samples for all ants indicate an overload (red). As a result, at the end of the phase, a few ants drop out—again with w.p. linear in $\gamma$; this time the drop out is 'permanent'.

The next phase is depicted in the right figure. Here the first sample indicate an overload (for all ants w.h.p.), the second a lack (green) for ants (for all ants w.h.p.). From here on no ant will join or leave the task for a polynomial number of steps (w.h.p.).

---

**Algorithm Ant**   (see Theorem 3.1)

**Input at round $t$:**   Noisy feedback $\mathbf{F}_j = (F_t^j)_{j \in [k]}$ for each task, learning parameter $\gamma \in [\gamma^*, 1/16]$
**Output at round $t$:** Assignment of the ant to a task $a_t \in \{\text{idle}, 1, 2, \ldots, k\}$

1: $c_d \leftarrow 19$ and $c_s \leftarrow 2\frac{1}{3}$
2: **for** At every step $t \geq 1$ **do**
3:   **if** $t \mod 2 = 1$ **then**
4:     $currentTask \leftarrow a_{t-1}$
5:     $\mathbf{s}_1 \leftarrow \mathbf{F}_t$   [receive feedback]
6:     **if** $currentTask \neq \text{idle}$, **then** $a_t \leftarrow \begin{cases} \text{idle} & \text{w.p. } c_s\gamma \\ currentTask & \text{otherwise} \end{cases}$
7:   **if** $t \mod 2 = 0$ **then**
8:     $\mathbf{s}_2 \leftarrow \mathbf{F}_t$   [receive feedback]
9:     **if** $currentTask = \text{idle}$ **then**
10:       $underloadedTasks \leftarrow \{j \in [k] \colon s_1^{(j)} = s_2^{(j)} = \texttt{lack}\}$
11:       $a_t \leftarrow \begin{cases} \text{Uniform}\,(underloadTasks) & \text{if } underloadTasks \neq \emptyset \\ \text{idle} & \text{otherwise} \end{cases}$
12:     **else**
13:       $a_t \leftarrow \begin{cases} \text{idle} & \text{w.p. } \gamma/c_d \text{ if } s_1^{currentTask} = s_2^{currentTask} = \texttt{overload} \\ currentTask & \text{otherwise} \end{cases}$
    **output** $a_t$

---

$O(nk)$. Finally, $R^{\approx}(t)$ can simply, by definition, be bounded by $O(\gamma \sum_{j \in [k]} d^{(j)})$ per time step; in fact this is, up to constants, tight as our lower bounds show.

In the remainder we give the formal statements. See Appendix A for the missing proofs.

In the following claims we consider a continuous interval of time steps $\mathcal{I}$ of length $t \leq n^4$ and we will analyze the dynamics of the system in this interval assuming an arbitrary initial configuration in step 0. For each ant $i$ we define $a_t^i$ to be the task the ant $i$ is assigned to at the end of round $t$, i.e., $a_t^i \in \{\text{idle}, 1, 2, \ldots, k\}$. In particular, $a_0^i$ denotes the initial assignment of ants. We start by defining the pair of 'good' events that occur w.h.p. during $\mathcal{I}$. For each task $j$ we define the *grey zone* $g_j = [-\gamma^* d^{(j)}, \gamma^* d^{(j)}]$ which intuitively corresponds to deficits, in which the feedback is very 'noisy'.

Let $\mathcal{E}_{feedback}$ be the event that for any step $t$ of the interval $\mathcal{I}$, any task $j$, and any ant $i$ the following holds. If $W_{t-1}^{(j)} - d^{(j)} \notin g_j$ then the feedback received by ant $i$ for task $j$ is *correct*, meaning that $F_t^j = \mathtt{lack}$ if $\Delta_{t-1}^{(j)} \geq \gamma^* d^{(j)}$ and $F_t^j = \mathtt{overload}$ if $\Delta_{t-1}^{(j)} \leq -\gamma^* d^{(j)}$.

Let $\mathcal{E}_{concentration}$ be the event that for any step $t$ of the interval $\mathcal{I}$, such that $t \bmod 2 = 0$ and for any task $j$ such that $W_t^{(j)} \geq (1-\gamma)d^{(j)}$ we have:

- $W_{t+1}^{(j)} \in [W_t^{(j)}(1 - 1.1c_s\gamma), W_t^{(j)}(1 - 0.9c_s\gamma)]$,

- $W_{t+2}^{(j)} \geq W_t^{(j)}(1 - \frac{3\gamma}{2c_d})$,

- if $W_t^{(j)} \geq d^{(j)}(1 + (1 + 1.2c_s)\gamma)$, then $W_{t+2}^{(j)} \leq W_t^{(j)}(1 - \frac{\gamma}{2c_d})$.

In the following we show, using standard techniques, that w.h.p. each of these events occurs.

**Claim 4.1.** *We have that:*

1. $\mathbb{P}[\mathcal{E}_{feedback}] \geq 1 - n^{-2}$ *and*

2. $\mathbb{P}[\mathcal{E}_{concentration}] \geq 1 - 2n^{-2}$,

In the following we condition on both of these events.

**Claim 4.2.** *Conditioned on events $\mathcal{E}_{feedback}$ and $\mathcal{E}_{concentration}$. For each task $j \in [k]$ there exists at most one time step $\tau_j \in \mathcal{I}$ such that $W_{\tau_j}^{(j)} \geq d^{(j)}(1+\gamma) > W_{\tau_j - 1}^{(j)}$ and if $\tau_j$ exists, then for any $\tau \in \mathcal{I}$ such that $\tau > \tau_j$ and $\tau \bmod 2 = 0$ we have:*
$$W_{\tau_j}^{(j)} \geq W_\tau^{(j)} \geq d^{(j)}(1+\gamma).$$

In the following, we will split the cost of the regret into three different more tractable costs: the regret induced by a significant overload $R^+(t)$, by being close to the demands $R^\approx(t)$ and by significant lacks $R^-(t)$. Let $c^+ = 1.2c_s$ and $c^- = 1 + 1.2c_s$ Let
$$r^+(t) = \sum_{j \in [k]} (W_t^{(j)} - (1+c^+\gamma)d^{(j)}) \cdot \mathbb{1}_{W_t^{(j)} - (1+c^+\gamma)d^{(j)} > 0}$$
$$r^-(t) = \sum_{j \in [k]} ((1-c^-\gamma)d^{(j)} - W_t^{(j)}) \cdot \mathbb{1}_{(1-c^-\gamma)d^{(j)} - W_t^{(j)} > 0}$$
$$r^\approx(t) = r(t) - r^+(t) - r^-(t).$$
Similarly, define
$$R^+(t) = \sum_{\tau \leq t} r^+(\tau), \qquad R^\approx(t) = \sum_{\tau \leq t} r^\approx(\tau), \qquad \text{and } R^-(t) = \sum_{\tau \leq t} r^-(\tau).$$
We have $R(t) = R^+(t) + R^\approx(t) + R^-(t)$.

We will simply bound $R^\approx(t)$ by using its definition resulting in $R^+(t) \leq \frac{2knc_d}{\gamma}$. Bounding the other two quantities is more involved; we will start by bounding $R^+(t)$.

Using Claim 4.2 we can upper bound the total regret related to the overload in all the tasks. The idea of the proof is that for a fixed task, we have that during the interval $\mathcal{I}$, there can only be at most one significant increase in the load. After such an increase in the load of that task $j$, the number of ants working on task $j$ will from there on drop every phase by roughly a factor $\gamma$ until the regret due to overload $(r^+(\tau))$ becomes zero. Integrating over the entire period results in a geometric series and summing over all tasks gives a regret of $R^+(t) \leq \frac{2nkc_d}{\gamma}$.

**Claim 4.3.** *Conditioned on $\mathcal{E}_{feedback}$ and $\mathcal{E}_{concentration}$. If $\gamma \leq 1/16$, then we have that: $R^+(t) \leq \frac{2nkc_d}{\gamma}$, and $r^+(\tau) > 0$ for at most $4kc_d \log n/\gamma$ different rounds of $\tau \in \mathcal{I}$.*

We say that all tasks are *saturated* in step $t$ if for all tasks $j \in [k]$ we have in step $t$ that $W_t^{(j)} \geq d^{(j)}(1-\gamma)$.

**Claim 4.4.** *Condition on $\mathcal{E}_{feedback}$ and $\mathcal{E}_{concentration}$. If in time step $t \in \mathcal{I}$ all tasks are saturated, then*

1. *in any step $t' > t$ such that $t' \mod 2 = 0$, all tasks are saturated,*
2. *in any step $t' \geq t$, $r^-(t') = 0$.*

In order to bound $R^-(t)$ we use the following potential functions, which measure the number of ants working on a significantly overloaded task and the number of significantly overloaded tasks, respectively. For any $t \in \mathbb{N}$ let
$$\Phi(t) = \sum_{j \in [k]} ((1+\gamma)d^{(j)} - W_{2t}^{(j)}) \cdot \mathbb{1}_{(1+\gamma)d^{(j)} > W_{2t}^{(j)}},$$
and
$$\Psi(t) = \sum_{j \in [k]} \mathbb{1}_{(1+\gamma)d^{(j)} > W_{2t}^{(j)}}.$$
Observe that functions $\Phi$ and $\Psi$ are defined for phase numbers not for steps. Phase number $t$ consists of two time steps $2t$ and $2t+1$. In the following we show that after 2 phases we have that either one of three events must occur: 1) the number of ants working on underloaded tasks increases significantly 2) the number of saturated tasks increases 3) all tasks are saturated.

**Claim 4.5.** *Condition on $\mathcal{E}_{feedback}$ and $\mathcal{E}_{concentration}$. We have that the functions $\Phi(t)$ and $\Psi(t)$ are non-increasing and moreover for an arbitrary $t$ such that in step $2t$ not all tasks are saturated and $2t \leq n^4 - 4$, we have that at least one of the following three events must happen:*

1. $\Phi(t+2) - \Phi(t) \leq -c\gamma n$ *for some constant $c > 0$*
2. $\Psi(t+2) - \Psi(t) \leq -1$,
3. *all tasks are saturated in step $2(t+2)$.*

From this we are able to derive a bound on $R^-(t)$ and by putting everything together we derive the desired bound on $R(t) = R^+(t) + R^\approx(t) + R^-(t)$.

**Lemma 4.6.** *Condition on $\mathcal{E}_{feedback}$ and $\mathcal{E}_{concentration}$. We have that*
$$R(t) \leq \frac{ckn}{\gamma} + c \sum_{j \in [k]} \gamma t d^{(j)},$$
*where $c$ is a constant. Moreover for all tasks $j \in [k]$ the absolute value of the deficit is bounded by $5\gamma d^{(j)}$ in all but $O(k \log n/\gamma)$ rounds.*

From Lemma 4.6 and Claim 4.1 we are able to establish our main theorem Theorem 3.1. Since the claims have been shown for any initial configuration we can simply cut the time horizon into intervals of length at most $n^4$ and use the bounds on the regret for each interval. See Appendix A for the proof.

## 5 Algorithm Precise Sigmoid - Theorem 3.2

In this section we present Algorithm Precise Sigmoid that is built on the foundations of Algorithm Ant and to takes advantage of the sigmoid noise to achieve an even better total regret than Algorithm Ant. Recall that all Algorithm Ant requires is that whenever the deficit is outside of the grey zone, then the probability to receive an incorrect feedback is at most $1/n^8$. Now if we choose a step size of roughly $\varepsilon\gamma$ then, due to the sigmoid noise, the probability of failure increases roughly to $1/n^{\varepsilon 8}$. However, by taking multiple samples instead of a single sample and by computing the median, we can amplify the probability that the median sample is correct to $1 - 1/n^8$. This means that the result of Theorem 3.1 applies with the same guarantees but at a much smaller step size, ultimately resulting in a much smaller regret.

More precisely, the modified version uses a step size of $\varepsilon\gamma/c_\chi$ and relies on synchronized phases of length $2m$, with $m = \lceil 2c_\chi/\varepsilon + 1 \rceil$. Each such phase consists of $m$ rounds during which the algorithm calculates the median of the feedback. This results in 2 samples per phase just as in Algorithm Ant and the rest of the algorithm is exactly the same as Algorithm Ant. Up to the different step size and the computation of the median, the proof is along the same lines as the proof of Algorithm Ant. See Appendix B for the proof.

---

**Algorithm Precise Sigmoid**      (see Theorem 3.2)

**Input at round $t$:**   Noisy feedback $\mathbf{F}_t = (F_t^j)_{j \in [k]}$ for each task, learning parameter $\gamma < 1/2$, precision parameter $\varepsilon < 1$.

**Output at round $t$:**  Assignment of the ant to a task $a_t \in \{\text{idle}, 1, 2, \ldots, k\}$.

1: $c_\chi \leftarrow 10$, $c_d \leftarrow 19$ and $c_s \leftarrow 2\frac{1}{3}$
2: $m \leftarrow \lceil 2c_\chi/\varepsilon + 1 \rceil$
3: **for** At every step $t \geq 1$ **do**
4:    $r \leftarrow t \bmod 2m$
5:    **if** $r = 1$ **then**
6:      $currentTask \leftarrow a_{t-1}$
7:    **if** $r \in [1, m]$ **then**
8:      $\mathbf{s}_r \leftarrow \mathbf{F}_t$    [receive feedback]
9:      $a_t \leftarrow a_{t-1}$
10:    **if** $r = m$ **then**
11:      $\hat{\mathbf{s}}_1 \leftarrow \text{median}(\mathbf{s}_{r'} : r' \in [1, m])$
12:      **if** $currentTask \neq \text{idle}$, **then** $a_t \leftarrow \begin{cases} \text{idle} & \text{w.p. } \varepsilon c_s \gamma / c_\chi \\ currentTask & \text{otherwise} \end{cases}$
13:    **if** $r \in [m+1, 2m-1] \cup \{0\}$ **then**
14:      $\mathbf{s}_r \leftarrow \mathbf{F}_t$    [receive feedback]
15:      $a_t \leftarrow a_{t-1}$
16:    **if** $r = 0$ **then**
17:      $\hat{\mathbf{s}}_2 \leftarrow \text{median}(\mathbf{s}_{r'} : r' \in [m+1, 2m-1] \cup \{0\})$
18:      **if** $currentTask = \text{idle}$ **then**
19:        $underloadedTasks \leftarrow \{j \in [k] : \hat{s}_1^{(j)} = \hat{s}_2^{(j)} = \texttt{lack}\}$
20:        $a_t \leftarrow \begin{cases} \text{Uniform}(underloadTasks) & \text{if } underloadTasks \neq \emptyset \\ \text{idle} & \text{otherwise} \end{cases}$
21:      **else**
22:        $a_t \leftarrow \begin{cases} \text{idle} & \text{w.p. } \gamma/(c_\chi c_d) \text{ if } \hat{s}_1^{currentTask} = \hat{s}_2^{currentTask} = \texttt{overload} \\ currentTask & \text{otherwise} \end{cases}$
    **output** $a_t$

---

## 6 Conclusion and Open Problems

We presented a simple proof-of-concept algorithm that achieves, given a suitable learning parameter $\gamma$, a fairly good assignment of ants. The algorithm is very resilient to noise, changes in demands, changes of the number of ants and even changes of the number of tasks. The algorithm embraces the seeming obstacle of full-synchronization (which we introduced to model the delay of information) to perform controlled oscillations. It would be interesting to see if variations of this algorithm also work in settings of less synchronization.

Moreover, it remains an open problem to understand if and by how much simple communication among the ants can help. In the adversarial setting for example, it is clear that even unlimited communication cannot help. This leads to the question of which other noise models would make sense to study and how to design experiments with real ants to gather more knowledge about the way noise affects the sensing.

## A  Missing Proofs of Section 4 and Proof of Theorem 3.1

*Proof of Claim 4.1.* The probability that any ant receives an incorrect feedback from task $j$ in step $t$ assuming that $|\Delta_t^{(j)}| \geq \gamma^* d_j$ is by the definition of $\gamma^*$ at most $n^{-8}$. By taking union bound over $n^4$ step, at most $n$ ants and at most $n$ tasks we get that the total probability of all 'bad' events is at most $n^{-2}$, hence $\mathbb{P}[\mathcal{E}_{feedback}] \geq 1 - n^{-2}$.

In the following we condition on $\mathcal{E}_{feedback}$. Fix any task $j$ and round $t$ such that $t \bmod 2 = 0$. We assume that $\gamma \leq 1/16$ and $W_t^{(j)} \geq (1-\gamma)d^{(j)}$. By Assumptions 2.1 we have $W_t^{(j)} \geq (1-\gamma)d^{(j)} \geq (1-\gamma)50000\log n/\gamma^2 \geq 120 \cdot \max\{c_s^2, c_d^2\} \cdot \log n/\gamma^2$.

Recall that $W_{t+1}^{(j)}$ denotes the random variable indicating the number of ants working in the second sample for the considered task $j$. We have $\mathbb{E}\left[W_{t+1}^{(j)}\right] = W_t^{(j)}(1 - c_s\gamma)$. By Chernoff bound we have for any $0 < \delta_2 < 1$:
$$\mathbb{P}\left[W_{t+1}^{(j)} \geq (1+\delta_2)\mathbb{E}\left[W_{t+1}^{(j)}\right]\right] \leq e^{-\delta_2^2 \mathbb{E}\left[W_{t+1}^{(j)}\right]/3}, \quad \mathbb{P}\left[W_{t+1}^{(j)} \leq (1-\delta_2)\mathbb{E}\left[W_{t+1}^{(j)}\right]\right] \leq e^{-\delta_2^2 \mathbb{E}\left[W_{t+1}^{(j)}\right]/2},$$
and by choosing $\delta_2 = \sqrt{\frac{24\log n}{\mathbb{E}[W_{t+1}^{(j)}]}} \leq 1$ each of these probabilities is upper bounded by $n^{-8}$. Moreover we have, since $\gamma \leq 1/16$, $c_s < 1/(2\gamma)$, and $w_1 \geq 120 c_s^2 \log n/\gamma^2$ that:
$$\delta_2 = \sqrt{\frac{24\log n}{\mathbb{E}[W_3]}} = \sqrt{\frac{24\log n}{w_1(1-c_s\gamma)}} \leq \frac{c_s\gamma}{10},$$
Hence $\mathbb{P}\left[W_{t+1}^{(j)} \in [W_t^{(j)}(1-1.1c_s\gamma), (1-0.9c_s\gamma)]\right] \geq 1 - n^{-6}$, because $W_t^{(j)}(1-c_s\gamma)(1-c_s\gamma/10) \geq W_t^{(j)}(1-1.1c_s\gamma)$ and $W_t^{(j)}(1-c_s\gamma)(1+c_s\gamma/10) \leq W_t^{(j)}(1-0.9c_s\gamma)$. Now if in step $t+2$ some ants decide to transition to state idle then each of them decides independently with probability $\gamma/c_d$ hence $\mathbb{E}\left[W_{t+2}^{(j)}\right] \geq W_t^{(j)}(1-\gamma/c_d)$. We have by Chernoff bound
$$\mathbb{P}\left[W_{t+2}^{(j)} \leq (1-\gamma/(2c_d))\mathbb{E}\left[W_{t+2}^{(j)}\right]\right] \leq \mathbb{P}\left[W_{t+2}^{(j)} \leq (1-\delta_{stay})\mathbb{E}\left[W_{t+2}^{(j)}\right]\right] \leq e^{-\delta_{stay}^2 \mathbb{E}[W_{t+2}^{(j)}]/2},$$
by choosing $\delta_{stay} = \sqrt{\frac{24\log n}{\mathbb{E}[W_{t+2}^{(j)}]}} \leq \gamma/(2c_d)$, we get that $\mathbb{P}\left[W_{t+2}^{(j)} \leq W_t^{(j)}(1 - 3\gamma/(2c_d))\right] \leq n^{-8}$ (because $(1-\gamma/c_d)(1-\gamma/(2c_d)) \geq (1-3\gamma/(2c_d))$).

Assume that $W_t^{(j)} \geq d^{(j)}(1 + (1+1.2c_s)\gamma)$. We have with probability $1 - n^{-8}$ that $W_{t+1}^{(j)} \geq d^{(j)}(1 + 2.2c_s\gamma)(1 - 0.9c_s\gamma) = d^{(j)}(1 + 2.1\gamma - 1.98\gamma^2) \geq d^{(j)}(1+\gamma)$, where in the last inequality we use the fact that $\gamma \leq 1/16$. Then, conditioned on $\mathcal{E}_{feedback}$, each ant working on task $j$ receives feedback `overload` in both samples in step $t$ and $t+1$. By the definition of the algorithm, Each such ant leaves the task in step $t+2$ with probability $\gamma/c_d$. Let random variable $W_{leave}$ denote the number of ants that leave task $j$ in step $t+2$. We can bound the value of this variable using Chernoff bound:
$$\mathbb{P}[W_{leave} \leq \mathbb{E}[W_{leave}]/2 | \mathcal{E}_{feedback}] \leq \mathbb{P}[W_{leave} \leq (1-\delta_{leave})\mathbb{E}[W_{leave}]] \leq e^{-\delta_{leave}^2 \mathbb{E}[W_{leave}]/2},$$
and by choosing $\delta_{leave} = \sqrt{\frac{18\log n}{\mathbb{E}[W_{leave}]}} \leq 1/2$, the inequality is true with probability at most $n^{-9}$. We observe that since $\mathbb{E}[W_{leave}|\mathcal{E}_{feedback}] = \gamma W_t^{(j)}/c_d$ we have that with probability of at most $n^{-9}$ at most $W_t^{(j)}\gamma/(2c_d)$ ants decide to leave task $j$ in step $t+2$. Thus $\mathbb{P}\left[W_{t+2}^{(j)} \geq W_t^{(j)}(1-\gamma/(2c_d))|\mathcal{E}_{feedback}\right] \leq n^{-9}$.

We showed that each bad event happens with probability at most $n^{-8}$. By the union bound over all ants (at most $n$), steps ($n^4$) and tasks (at most $k$) and 4 different bad events we get that with probability at least $1 - n^{-2}$ no bad event happens in the considered interval. This shows that $\mathbb{P}[\mathcal{E}_{concentration}|\mathcal{E}_{feedback}] \geq 1 - n^2$ but since $\mathbb{P}[\mathcal{E}_{feedback}] \geq 1 - n^{-2}$ we get $\mathbb{P}[\mathcal{E}_{concentration}] \geq \mathbb{P}[\mathcal{E}_{concentration}|\mathcal{E}_{feedback}]\mathbb{P}[\mathcal{E}_{feedback}] \geq (1 - n^{-2})^2 \geq 1 - 2n^{-2}$.
□

*Proof of Claim 4.2.* First observe that, since the number of ants working on a task increases only in steps $\tau$ such that $\tau \bmod 2 = 0$ then if there exists $\tau$ such that $W_\tau^{(j)} \geq d_j(1 + \gamma)$, then the smallest such $\tau$ (call it $\tau_j$) must satisfy $\tau_j \bmod 2 = 0$. Fix a task $j$ and assume that $\tau_j$ exists. We observe that conditioned on $\mathcal{E}_{feedback}$ all idle ants receive feedback `overload` in the first sample for task $j$ in step $\tau_j$. Hence no ant can join task $j$ in this phase (recall that in order to join, both samples of the phase need to show `lack`). This shows that whenever the load is at least $d_j(1+\gamma)$ in the first step of a phase then it cannot be larger in the first step of the next phase.

Second, observe that if $W_{\tau_j}^{(j)} \leq d^{(j)}(1 + (0.9c_s - 1)\gamma)$ then conditioning on event $\mathcal{E}_{concentration}$, the second sample in the phase starting in step $\tau_j$ satisfies $W_{\tau_j+1}^{(j)} \leq W_{\tau_j}^{(j)}(1 - 0.9c_s\gamma) \leq d^{(j)}(1 - \gamma)$ thus conditioning on $\mathcal{E}_{feedback}$, all ants receive feedback `lack` in the second sample thus no ant can leave this task in this phase. These two claims show the existence of a *stable zone* $[d^{(j)}(1+\gamma), d^{(j)}(1+(0.9c_s-1)\gamma)]$ (which is nonempty because $c_s > 2$), where the number of ants in the task can neither increase nor decrease (if we look only at steps $\tau$ such that $\tau \bmod 2 = 0$).

Observe moreover that if $W_{\tau_j}^{(j)} \geq d^{(j)}(1+(0.9c_s - 1)\gamma)$, then conditioning on event $\mathcal{E}_{concentration}$ always at least $W_\tau^{(j)}(1 - 3\gamma/(2c_d))$ ants remain in this task. We have then $W_{\tau_j+2}^{(j)} \geq W_{\tau_j}^{(j)}(1 - 3\gamma/(2c_d)) \geq d^{(j)}(1 + (0.9c_s - 1)c_s)(1 - 3\gamma/(2c_d))$. The last expression is lower bounded by $d^{(j)}(1 + \gamma)$ if $c_s \geq 20/9 + 2/(c_d - 1)$. Observe that with our choice of $c_s$ and $c_d$ this holds. This shows that it is not possible to 'jump over' the stable zone.

Now we can inductively consider steps $\tau_j + 2, \tau_j + 4, \ldots$ and assuming that $\tau_j + 2i, \tau_j + 2(i+1) \in \mathcal{I}$ and conditioned on $\mathcal{E}_{concentration}$ and $\mathcal{E}_{feedback}$ we have that $W_{\tau_j+2i}^{(j)} \geq W_{\tau_j+2(i+1)}^{(j)} \geq d^{(j)}(1+\gamma)$.
□

*Proof of Claim 4.3.* Fix any task $j \in [k]$. Let us define the regret due to overload with respect to task $j$:
$$r_j^+(\tau) = (W_\tau^{(j)} - (1+c^+\gamma)d^{(j)}) \cdot \mathbb{1}_{W_\tau^{(j)} - (1+c^+\gamma)d^{(j)} > 0}$$
and
$$R_j^+(t) = \sum_{\tau \leq t} r_j^+(\tau).$$

Clearly $R^+(t) = \sum_{j \in [k]} R_j^+(t)$.

Let us define the smallest step $\tau^*$ such that $\tau^*(\bmod 2) = 0$ and $W_{\tau_j}^{(j)} \geq d^{(j)}(1 + c^+\gamma)$. If such a step $\tau^*$ does not exist in interval $\mathcal{I}$ then $R_j^+(t) = 0$.

In the opposite case we can define a sequence of steps $t_l = \tau^* + 2 \cdot l$. Conditioned on $\mathcal{E}_{concentration}$ if for some $l \geq 1$, we have $W_{t_l}^{(j)} \geq d^{(j)}(1+c^+\gamma)$, then $W_{t_{l+1}}^{(j)} \leq W_{t_l}^{(j)}(1 - \gamma/(2c_d)) \leq W_{\tau_j}^{(j)}(1 - \gamma/(2c_d))^l$. So for some $l^* \leq \log_{1/(1-\gamma/(2c_d))} n$ we must have that $W_{t_{l^*}}^{(j)} \leq d^{(j)}(1 + c^+\gamma)$. And by Claim 4.2 in any step $\tau \geq t_{l^*}$ we have $W_\tau^{(j)} \leq W_{t_{l^*}}^{(j)} \leq d^{(j)}(1 + c^+\gamma)$. Observe that for any phase consisting of two time steps $\tau, \tau+1$ such that $\tau(\bmod 2) = 0$ then in step $\tau + 1$ ants only leave the task hence $W_{\tau+1}^{(j)} \leq W_\tau^{(j)}$ and $W_{\tau+1}^{(j)} \leq W_\tau^{(j)}$ thus $r_j^+(\tau) + r_j^+(\tau+1) \leq 2r_j^+(\tau)$. We have:
$$R_j^+(t) = \sum_{\tau \leq t} r_j^+(\tau) = \sum_{\tau < \tau^*} r_j^+(\tau) + \sum_{\tau^* \leq \tau < t_{l^*}} r_j^+(\tau) + \sum_{\tau \geq t_{l^*}} r_j^+(\tau) \leq \sum_{\tau^* \leq \tau < t_{l^*}} r_j^+(\tau)$$
$$\leq 2\sum_{l=0}^{l^*-1} r_j^+(t_l) \leq 2\sum_{l=0}^{l^*-1} n \cdot (1 - \gamma/(2c_d))^l \leq 2n\sum_{l=0}^{\infty}(1 - \gamma/(2c_d))^l \leq 2nc_d/\gamma$$

Finally $R^+(t) = \sum_{j \in [k]} R_j^+(t) \leq 2nkc_d/\gamma$. Observe that $r^+(\tau) > 0$ only for at most
$$2kl^* = 2k\log_{1/(1-\gamma/(2c_d))} n = \frac{2k \ln n}{\ln \frac{1}{1-\gamma/(2c_d)}} \leq 4kc_d \log n/\gamma$$
steps in the considered interval, where we in the last inequality we used that $\ln x \geq (x-1)/x$. □

*Proof of Claim 4.4.* To prove the first claim take any task $j \in [k]$ and observe that if $W_t^{(j)} \geq d^{(j)}(1+\gamma)$ then it remains at least $d^{(j)}(1+\gamma)$ in all even time steps by Claim 4.2. On the other hand if $W_t^{(j)} \in [d^{(j)}(1-\gamma), d^{(j)}(1+\gamma)]$ then in the second sample in this phase we have by $\mathcal{E}_{concentration}$ at most $d^{(j)}(1+\gamma)(1-0.9c_s\gamma) \leq d^{(j)}(1-(0.9c_s-1)\gamma)$ and since $0.9c_s \geq 2$, the last expression is upper bounded by $d^{(j)}(1-\gamma)$ hence by $\mathcal{E}_{feedback}$ all the ants receive feedback lack in the second sample in the considered phase. Hence no ant can leave the task in this phase. Thus we also have that in step $t+2$ all tasks are saturated.

We observe that if for a task $j$, $W_t^{(j)} \geq d^{(j)}(1-\gamma)$, then conditioned on $\mathcal{E}_{concentration}$, in the second sample in this phase the number of workers in this task is at least $d^{(j)}(1-\gamma)(1-1.1c_s\gamma) > d^{(j)}(1-c^-\gamma)$ combined with the fact that no ant can leave the task at the end of the phase this shows the second claim. □

*Proof of Claim 4.5.* First, we observe that $\Phi(\cdot)$ and $\Psi(\cdot)$ are non-increasing, because due to $\mathcal{E}_{feedback}$ and Claim 4.2, no ant can join a task $j$ that has load of at least $d^{(j)}(1+\gamma)$ at the first step of a phase.

We distinguish between two cases depending on $\iota_t$: the number of idle ants at time $2t$.

1. Suppose $\iota_t \geq \frac{\gamma n}{64}$. We know that not all tasks are saturated hence there exists in at the beginning of step $2t+1$ at least one task $j$ with at most $d^{(j)}(1-\gamma)$ workers. By $\mathcal{E}_{feedback}$ all idle ants receive feedback lack from task $j$ at the beginning of step $2(t+1)$. Hence, by the definition of the algorithm, all the idle ants join some tasks in step $2(t+1)$. By $\mathcal{E}_{feedback}$ all idle ants will join tasks $\{j': W_{2t}^{(j')} < d^{(j')}(1+\gamma)\}$. Again, there are two cases, either there is now at least one task $j$ for which load is now above $d^{(j)}(1+\gamma)$ and we have that $\Psi(t+2) - \Psi(t) \leq \Psi(t+1) - \Psi(t) \leq -1$, due to monotonicity of $\Psi(\cdot)$. Otherwise, $\Phi(t+2) - \Phi(t) \leq \Phi(t+1) - \Phi(t) \leq -\frac{\gamma n}{64}$.

2. Now suppose $\iota_t < \frac{\gamma n}{64}$. Since we assume (Assumptions 2.1) that $\sum_{j \in [k]} d^{(j)} \leq n/2$. Moreover, we assume $\gamma \leq 1/16$ and $c_s = 2\frac{1}{3}$ and hence $\sum_{j \in [k]}(1+(1+1.2c_s)\gamma)d^{(j)} \leq (1+\frac{1}{4})\frac{n}{2} = (1-\frac{3}{8})n$ and we get that the sum of ants working on tasks $j$ with $W_t^{(j)} \geq (1+(1+1.2c_s))\gamma d^{(j)}$ is at least
$$\mathcal{W} = \sum_{j \in [k]} W_t^{(j)} \mathbb{1}_{W_t^{(j)} \geq (1+(1+1.2c_s))\gamma d^{(j)}} \geq n - \frac{\gamma n}{64} - \left(1 - \frac{3}{8}\right)n \geq \frac{n}{4}.$$
By definition of the algorithm, after one phase each of these $\mathcal{W}$ many ants leaves w.p. $\gamma/c_d$ and by $\mathcal{E}_{concentration}$, at least $\gamma\mathcal{W}/(2c_d)$ resulting in $\iota_{t+1} \geq \frac{\gamma n}{4c_d}$ idle ants at the beginning of the next phase. From here on the same argument as above holds and we will have that either all tasks are saturated in step $2(t+1)$ or $\Psi(t+2) - \Psi(t) \leq -1$ or $\Phi(t+2) - \Phi(t) \leq -\frac{\gamma n}{4c_d}$.

□

*Proof of Lemma 4.6.* The proof consists of three parts.

First observe that for any round $\tau$, the regret in that round is trivially bounded by:
$$r(\tau) \leq \sum_{j \in [k]} |X_\tau^j - d^{(j)}| \leq \sum_{j \in [k]} X_\tau^j + \sum_{j \in [k]} d^{(j)} \leq 2n \tag{1}$$

We know by Claim 4.3 that:
$$R^+(t) \leq \frac{2knc_d}{\gamma}. \tag{2}$$

Secondly, by the definition of $R^{\approx}(t)$:
$$R^{\approx}(t) \le \max\{c^+, c^-\} \sum_{j \in [k]} \gamma t d^{(j)}. \tag{3}$$
Finally we need to show that, for some constant $c_2$:
$$R^-(t) \le \frac{c_2 k n}{\gamma}. \tag{4}$$
To prove this we use the potential argument from Claim 4.5. By Claim 4.5 we have that for any step $\tau$ such that $\tau \mod 2 = 0$ if $\Phi(\tau) = 0$ and $\Psi(\tau) = 0$ then all the tasks are saturated which in turn implies by Claim 4.4 that $r^-(\tau') = 0$ for all $\tau' \ge \tau$. Observe that $\Psi(0) \le k$ and $\Phi(0) \le 2n$ and by Claim 4.5 from the monotonicity of $\Psi$ and $\Phi$ and the rate at which these potentials decrease there can be at most $2\left(\frac{2n}{c\gamma n} + k\right) = O(k + \frac{1}{\gamma})$ steps in which not all tasks are saturated. The regret per round is, by (1), bounded by $2n$, which yields (4).

We also showed that in at most $k + \frac{1}{\gamma}$ steps $r^-(t) > 0$. By Claim 4.3 in at most $O(k \log n/\gamma)$ steps we have $r^+(t) > 0$. Thus in all but $O(k \log n/\gamma)$ steps in all tasks $j \in [k]$ the absolute value of the deficit is bounded by $5\gamma d^{(j)}$.

$\square$

*Proof of Theorem 3.1.* From Lemma 4.6 it follows that the total regret is w.h.p. during the interval $\mathcal{I}$ is, by Union bound, w.p. $p \ge 1 - 1/n$ bounded by
$$\frac{ckn}{\gamma} + c^{-1} \sum_{j \in [k]} \gamma t d^{(j)}.$$

In the case that the high probability event fails, we simply use the following crude bound: The regret in any interval of length $n^4$ is bounded by $2n^5$, by (1).

Now consider an arbitrary time step $t \in \mathbb{N}$ (possibly super-exponential in $n$). Let $\ell = \lfloor t/n^4 \rfloor$ denote the number of complete intervals that are contained in the first $t$ time steps. Let $B$ denote the number of complete intervals among the first $\ell$ in which the regret is exceeds the upper bound in the statement of Lemma 4.6. Observe that $\mathbb{E}[B] \le (1-p)\ell \le \frac{t}{n^6}$. We have
$$B \sim \text{Binomial}(\ell, \mathbb{P}[\mathcal{E}_{concentration} \cap \mathcal{E}_{feedback}]) \precsim \text{Binomial}(\ell, n^{-1}),$$
where we use the symbol $\sim$ to denote "distributed as" and the symbol $\precsim$ to express stochastic domination. Hence, by Markov inequality, $\mathbb{P}[B \ge \frac{t}{n^5}] \le \mathbb{P}[B \ge n\mathbb{E}[B]] \le 1/n$ and therefore, by Union bound, we have that w.h.p.

$$R(t) \le B 2n^5 + (\ell - X)\left(\frac{ckn}{\gamma} + c^{-1}\gamma \sum_{j \in [k]} n^4 d^{(j)}\right) + \left(\frac{ckn}{\gamma} + c^{-1}\gamma \sum_{j \in [k]} (t - \ell n^4) d^{(j)}\right)$$
$$\le 2t + (1 + \ell)\frac{ckn}{\gamma} + c^{-1}\gamma \sum_{j \in [k]} t d^{(j)} \le 2t + \left(1 + \frac{t}{n^4}\right)\frac{ckn}{\gamma} + c^{-1}\gamma \sum_{j \in [k]} t d^{(j)}$$
$$\le \frac{ckn}{\gamma} + c^{-1}\gamma \sum_{j \in [k]} t d^{(j)} + 3t.$$

This concludes the proof.

$\square$

# B  Sigmoid Noise - Missing Proofs of Section 3.3

*Proof of Theorem 3.3.* Consider the setting where $k = O(1)$. We choose the constant $c$ in the statement of Theorem 3.3 such that each $A_i$ uses at most $\left\lfloor \log\left(\frac{1}{16\sqrt{\varepsilon}}\right) \right\rfloor$ bits of memory.

This implies that the corresponding state-machine has at most $s = 2^{\lfloor \log\left(\frac{1}{16\sqrt{\varepsilon}}\right) \rfloor} = \frac{1}{16\sqrt{\varepsilon}}$ states. Furthermore, we assume that all transition probabilities are either 0 or lower bounded by some $p \in (0,1)$, which can be a function of $\varepsilon$, but not of $n$.

Consider an interval of length $s$ and remark that $s$ is a constant since $\varepsilon$ is a constant. In the remainder we will show that there exists for each task $j \in [k]$ w.p. at least $1 - e^{-\Omega(n)}$ a time step $\tau$ such that
$$|\Delta_\tau^{(j)}| \geq 2s\varepsilon\gamma^* \sum_{j \in [k]} d^{(j)}. \tag{5}$$

Hence, the regret in the interval of length $s$ is at least $2s\varepsilon\gamma^* \sum_{j \in [k]} d^{(j)}$ for all tasks $j \in [k]$ with overwhelming probability, by taking Union bound. From this we easily get that for any $t$ large enough:
$$R(t) \geq \left(\varepsilon\gamma^* \sum_{j \in [k]} d^{(j)}\right) t,$$
due to the following reasoning. Consider $t'$ many time intervals of length $s$. Let $X_{t'}$ be the number of time intervals in which the regret is larger than $2s\varepsilon\gamma^* \sum_{j \in [k]} d^{(j)}$. We have that $\mathbb{E}[X_{t'}] = t'e^{-\Omega(n)}$. Thus, by Chernoff bounds (the random choices made in each interval are independent),
$$\mathbb{P}\left[X_{t'} \geq s\varepsilon\gamma^* \sum_{j \in [k]} d^{(j)} t'\right] \leq \mathbb{P}\left[X_{t'} \geq \frac{\mathbb{E}[X_{t'}]}{2}\right] \leq e^{-t'\Omega(n)}.$$
Thus, by Union bound, $\sum_{t'=1}^{\infty} \mathbb{P}[X_{t'}] \leq e^{-\Omega(n)+1}$. It remains to show (5).

We set $d^{(j)} = \sqrt{n}$, for $j \in [k]$. Let $\gamma' = 2s\varepsilon\gamma^*$. Recall that for any task $j \in [k]$:
$$\frac{1}{1 + e^{\lambda\gamma^* d^{(j)}}} = 1 - s(\gamma^* d^{(j)}) = s(-\gamma^* d^{(j)}) = \frac{1}{n^8}.$$
Assuming that the load of a task $j$ is in $[(1-\gamma')d^{(j)}, (1+\gamma')d^{(j)}]$, we have that the probability of receiving incorrect feedback for each ant is at least, using the symmetry of the sigmoid function,
$$q = \min\{s(x) \colon x \in [-\gamma' d^{(j)}, \gamma' d^{(j)}]\} = s(-\gamma' d^{(j)}) = \frac{1}{1 + e^{\lambda s\varepsilon\gamma^* d^{(j)}}} \geq \frac{1}{e} \frac{1}{e^{\lambda 2s\varepsilon\gamma^* d^{(j)}}}$$
$$= \frac{1}{e}\left(\frac{1}{e^{\lambda\gamma^* d^{(j)}}}\right)^{2s\varepsilon} \geq \frac{1}{e}\left(\frac{1}{1+e^{\lambda\gamma^* d^{(j)}}}\right)^{2s\varepsilon} \geq \frac{1}{en^{16s\varepsilon}}.$$

By the Pigeonhole principle and Assumptions 2.2, there must exist a task $i$ such that there must exist a sequence of feedback signals (possibly different for each ant) such that $n/k$ ants joins task $i$. W.l.o.g. we assume that the probability of receiving an incorrect feedback signal for task $i$ in every round throughout the interval is at least $q$; if it is less than this value, then this implies that the regret is already of order $\gamma' d^{(j)}$. Thus, the probability for every ant to join is at least $p' = (p \cdot q)^s \geq (p \cdot \frac{1}{en^{16s\varepsilon}})^s = p^{16s\varepsilon} n^{-16/(16)^2} e^s$.

Let $Y_i$ denote the number of ants joining task $i$ in the interval of length $s$. Note that $Y_i$ is binomially distributed with parameters $n/k$ and $p'$. We have $\mathbb{E}[Y_i] \geq \frac{n}{k}p' \geq 2n^{2/3}$. By Chernoff inequality with $\delta = 1/2$ we get
$$\mathbb{P}\left[Y_i \geq n^{2/3}\right] \leq \mathbb{P}\left[Y_i \geq \frac{\mathbb{E}[Y_i]}{2}\right] \leq \exp\left(-\frac{2n^{2/3}}{12}\right) = e^{-\Omega(n)}. \tag{6}$$

Summing over all $k$ tasks gives that

$$\sum_{\tau=1}^{s} \sum_{j \in [k]} |d^{(j)} - W_\tau^{(j)}| \geq \sum_{j \in [k]} \min\{Y_i - d^{(j)}, 0\} e^{-\Omega(n)} \geq n^{2/3} e^{-\Omega(n)}$$
$$\geq \sum_{j \in [k]} 2\varepsilon \gamma^* d^{(j)},$$

This yields (5) and completes the proof. □

*Proof of Theorem 3.2.* Recall that $s(-\gamma^* d^{(j)}) \leq 1/n^8$. For any task $j \in [k]$
$$\frac{1}{1 + e^{\lambda \gamma d^{(j)}}} = s(-\gamma d^{(j)}) \leq s(-\gamma^* d^{(j)}) = \frac{1}{n^8}.$$
Moreover,
$$\frac{1/e}{e^{\lambda \gamma d^{(j)}}} \leq \frac{1}{1 + e^{\lambda \gamma d^{(j)}}}$$
and hence, by putting both together, we get
$$\frac{1}{e^{\lambda \gamma d^{(j)}}} \leq \frac{e}{n^8}.$$
From this we get that for a step size of $\gamma' = \varepsilon \gamma / c_\chi$ the probability $p$ that the sample is incorrect, is at most,
$$p = s(-\gamma' d^{(j)}) = \frac{1}{1 + e^{\lambda \varepsilon \gamma d^{(j)}/c_\chi}} \leq \frac{1}{e^{\lambda \varepsilon \gamma d^{(j)}/c_\chi}} = \left(\frac{e}{n^8}\right)^{\varepsilon/c_\chi}$$

Thus, the event $Y$ that the median computed for a single round for a single ant is incorrect is, by Theorem E.3 with $\alpha = 1/2$, bounded by
$$\mathbb{P}\left[Y \geq \frac{m}{2}\right] \leq \left((2p)^{1/2} (2)^{1/2}\right)^m \leq 2^m \left(\left(\frac{e}{n^8}\right)^{\varepsilon/(2c_\chi)}\right)^m \leq \frac{1}{n^8}.$$

We observe that these are exactly the same guarantees that Algorithm Ant requires and hence we get from Theorem 3.1 using a step size of $\gamma'$ instead of $\gamma$ a regret of
$$R(t) \leq c\frac{nk}{\gamma'} + \left(\varepsilon \gamma \sum_{j \in [k]} d^{(j)} + O(1)\right) \cdot t,$$
for some constant $c$. We note that it will take longer to converge to right value as the step size is smaller and the rounds are longer. □

## C Adversarial Noise

We first give the intuition behind Algorithm Precise Adversarial. Again, the algorithm builds on Algorithm Ant. This time, every phase consists of two sub-phases. In the first sub-phase, the algorithm takes samples spaced roughly $\varepsilon \gamma / 32$ apart. Each ant then remembers the task $j$ (including the idle state) it executed when it received for the first time the feedback `lack` (during the first-sub phase the feedback).

Throughout the second sub-phase, the ant will execute task $j$.

The idea behind this is that when an ant received for the first time the feedback `lack`, the deficit must have been close to 0. Hence, taking many samples in this region yields a small regret.

We now give the sketch of Theorem 3.6.

*Proof of Theorem 3.6.* Algorithm Precise Adversarial is variation of Algorithm Ant with the difference that $O(1/\varepsilon)$ samples are taken in each phase instead of just 2. The main ideas are the same; in particular, if all samples indicate an overload, then after the phase, the number of ants reduces by a factor of roughly $\gamma$.

Each phase consists of two sub-phases, the first one covering $O(1/\varepsilon)$ rounds that is used to find the first round $r_{min}$ in the sub-phase where the the feedback received by the ant switched from `overload` to `lack`.

---

**Algorithm Precise Adversarial**     (see Theorem 3.6)

**Input at round $t$:**   Noisy feedback $\mathbf{F}_t = (F_t^j)_{j \in [k]}$ for each task, learning parameter $\gamma \in [\gamma^*, 1/16]$, precision parameter $\varepsilon < 1$

**Output at round $t$:**   Assignment of the ant to a task $a_t \in \{\text{idle}, 1, 2, \ldots, k\}$

$r_1 \leftarrow \frac{32}{\varepsilon}$, $r_2 \leftarrow 4r_1$
**for** At every step $t \geq 1$ **do**
  $r \leftarrow t \bmod (r_1 + r_2)$
  **if** $r = 1$ **then**
    $\mathbf{s}_1 \leftarrow \mathbf{F}_t$     [receive feedback]
    $currentTask \leftarrow a_{t-1}$
    $t' \leftarrow t$
  **if** $r \in [2, r_1)$ **then**
    $\mathbf{s}_r \leftarrow \mathbf{F}_t$     [receive feedback]
    **if** $currentTask \neq \text{idle}$, **then** $a_t \leftarrow \begin{cases} \text{idle} & \text{w.p. } \frac{\varepsilon\gamma}{32} \\ currentTask & \text{otherwise} \end{cases}$
  **if** $r = r_1$ **then**
    $r_{\min} \leftarrow \min\{r' < r_1 \colon s_{r'}^{(j)} = \texttt{lack} \text{ or } r' = r_1\}$
    $\mathbf{s}_r \leftarrow \mathbf{F}_t$     [receive feedback]
    **if** $currentTask \neq \text{idle}$, **then** $a_t \leftarrow \begin{cases} \text{idle} & \text{if } a_{t'+r_{\min}-1} = \text{idle} \\ currentTask & \text{otherwise} \end{cases}$
  **if** $r \in [r_1+1, r_1+r_2-1]$ **then**
    $a_t \leftarrow a_{r_{\min}}$
  **if** $r = 0$ **then**
    $\mathbf{s}_0 \leftarrow \mathbf{F}_t$     [receive feedback]
    **if** $currentTask = \text{idle}$ **then**
      $underloadedTasks \leftarrow \{j \in [k] \colon s_0^{(j)} = s_1^{(j)} = \cdots = s_{r_1+r_2-1}^{(j)} = \texttt{lack}\}$
      $a_t \leftarrow \begin{cases} \text{Uniform}(underloadTasks) & \text{if } underloadTasks \neq \emptyset \\ \text{idle} & \text{otherwise} \end{cases}$
    **else**
      $a_t \leftarrow \begin{cases} \text{idle} & \text{w.p. } \frac{\varepsilon\gamma}{32} \text{ if } s_0^{currentTask} = s_1^{currentTask} = \cdots = s_{r_1+r_2-1}^{currentTask} = \texttt{overload} \\ currentTask & \text{otherwise} \end{cases}$
  **output** $a_t$

---

Then in the second sub-phase of length $r_2 = O(1/\varepsilon)$ samples are taken as follows: if an ant was working in round $r_{\min}$ (which may differ from ant to ant) then it will work throughout $r_2$; and otherwise it will be idle throughout $r_2$.

We now argue that w.h.p. the average regret per phase is small for the case that load is not in a zone where it will either decrease or increase w.r.t. to the beginning of the following phase. Let $c = 1/16$. Fix an arbitrary task $j \in [k]$. Assume that at the beginning of the phase we had $W_t^{(j)} \leq (1 + 3\gamma)d^{(j)}$ since otherwise w.h.p. all ants working on $j$ will see an overload. Moreover, we assume that at the beginning of the phase it must hold that $W_t^{(j)} \geq (1 - 2\gamma)d^{(j)}$ since otherwise, the resource is underloaded.

Therefore, throughout one can show that w.h.p. the first sub-phase we have $|\Delta_t^{(j)}| \leq (1 + 4\gamma)d^{(j)}$. Furthermore, we have w.h.p. that throughout the second sub-phase it holds that $|\Delta_t^{(j)}| \leq (1 + c\varepsilon\gamma)d^{(j)}$.

By definition,
$$r_2 = 4r_1$$

This implies that w.h.p. the total regret during the phase is at most:
$$(4r_1 + c\varepsilon r_2) \sum_{j\in[k]} d^{(j)}\gamma \leq (1+\varepsilon)(r_1 + r_2)\gamma \sum_{j\in[k]} d^{(j)},$$
where the last inequality holds since $r_2 = 4r_1 \geq \frac{4-(1+\varepsilon)}{1+\varepsilon-c\varepsilon}r_1$. Similar as in Theorem 3.1, we can bounded the regret in case that the high probability events did not hold. Summing over all phases yields the desired result.

□

We now give the proof of Theorem 3.5.

*Proof of Theorem 3.5.* We argue that an algorithm cannot distinguish between two different demands. We define two demand vectors. Let $\mathbf{d} = (n/(2k), n/(2k), \ldots, n/(2k))$ and $\mathbf{d}' = (n/(2k) - 2\tau, n/(2k) - 2\tau, \ldots, n/(2k) - 2\tau)$, where $\tau = (1-o(1))\gamma^{ad}n/(2k)$ is the threshold parameter defined in Section 2.2. We now define the responses of the adversary

For the case of $\mathbf{d}$ we define the responses
$$F_i^t(j) = \begin{cases} 1 & \Delta_{t-1}^{(j)} \geq -\gamma^{ad}d^{(j)} \\ 0 & \Delta_{t-1}^{(j)} < -\gamma^{ad}d^{(j)} \end{cases}.$$

Similarly, for the case of $\mathbf{d}'$ we define the responses
$$F_i^t(j) = \begin{cases} 1 & \Delta_{t-1}^{(j)} \geq \gamma^{ad}d^{(j)} \\ 0 & \Delta_{t-1}^{(j)} < \gamma^{ad}d^{(j)} \end{cases}.$$

It can easily be verified that the feedback in both cases is identical for any load distribution.

We apply Yao's principle Theorem E.1: With probability $1/2$ we choose the demand vector $\mathbf{d}$ and otherwise $\mathbf{d}'$; we assume that the algorithm knows this input distribution. The cost of the algorithm is simply the demand.

Consider the best deterministic algorithm $A$ on this input distribution. We have

$$\mathbb{E}[R(t)] \geq \sum_{t' \leq t} \sum_{j\in[k]} \left(\frac{1}{2}\left|W_\tau^{(j)} - \frac{n}{2k}\right| + \frac{1}{2}\left|W_\tau^{(j)} - \frac{n}{2k} + 2\tau\right|\right) \geq tk\tau. \quad (7)$$

By Theorem E.1 any randomized algorithm has at least an regret of $tk\tau = (1-o(1))t\gamma^* \sum_{j\in[k]} d^{(j)}$ over the course of $t$ time steps. This concludes the proof.

□

## D  Trivial Algorithm

In this section we consider for the sake of completeness the following trivial algorithm. Each ant that is currently not working and receives the feedback that there is lack of ants a task, joins one of those tasks (chosen u.a.r.); if the ant is already working at a task, then it continues working until it receives feedback that there is an overload. We will analyze this algorithm in two models:

1. **Sequential model**: here exactly one ant is chosen u.a.r. and receives the feedback of the round before.

2. **Synchronous model**: all ants receive feedback simultaneously. See Section 2 for details.

## D.1 Sequential Model

In this section we show that the trivial algorithm performs reasonably well in this model. The reason is that once some fixed task $j$ is slightly overloaded (by a constant factor), then all other ants will see this and refrain from joining. To make this more concrete, we can again use the critical value $\gamma^*$. Once the overload (or underload) is at least $\gamma^* d^{(j)}$, each following ant—for the next $n^4$ time steps, by the same argument as in Claim 4.1—will w.h.p. receive the correct feedback and will refrain from joining (or leaving).

It is also not too hard to see that the obtained regret is up to a constant factor intrinsic to this algorithm (and any other constant memory sequential algorithm): The proof follows from a very similar algorithm as Theorem 3.3. Consider the setting with one task with demand $d^{(1)} = \sqrt{n}$. Suppose the overload is less than $\gamma^* d^{(1)}/20$, then the probability for an idle ant to join the task is $s(-\gamma^* d^{(1)}/20)$ which is roughly $1/(en^{8/20})$. Hence, the expected increase in the number of ants working (assuming ants are chosen u.a.r.) is at least:

$$\mathbb{E}\left[W_t^{(j)} - W_{t-1}^{(j)} \mid \mathcal{F}_{t-1}\right] \geq \frac{(n - W_{t-1}^{(j)})}{n} \frac{1}{en^{8/20}} - \frac{W_{t-1}^{(j)}}{n} \geq \frac{1}{en^{8/20}} - \frac{2W_{t-1}^{(j)}}{n} \geq \frac{1}{2en^{8/20}},$$

where we used our assumption that $\gamma^*$ is a constant and we generously assumed that every working ant leaves the task. This positive drift, shows that as long the overload is less than $\gamma^* d^{(1)}/20$, the system will slowly converge to an overload of $\gamma^* d^{(1)}/20$. From here one can use standard techniques to conclude and we note that the proof generalizes to any constant number of jobs $k$. Therefore, the regret is of size $\lim_{t \to \infty} \frac{R(t)}{t} = \Theta(\gamma^* \sum_{j \in [k]} d^{(j)})$, which asymptotically matches the optimal regret in the synchronous setting (Theorem 3.1 and Theorem 3.3).

## D.2 Synchronous Model

In this section we argue that the trivial algorithm fails to achieve a good task allocation in the synchronous model. Concretely, we show that the trivial algorithm does not converge in $e^{\Omega(n)}$ steps. Consider the example with one task having a demand $d^{(1)} = n/4$ and assume $\gamma^*$ to be some constant smaller than $1/2$. Assume that the initial load is 0, then with overwhelming probability all ants will receive `lack` and join. In the following step all ants will receive with overwhelming probability the feedback `overload` and all leave. Union bound, these oscillations continue for at least $e^{\Omega(n)}$ steps.

# E Auxiliary Tools

Given a problem, let $\mathcal{X}$ be the set of inputs and $\mathcal{A}$ be the set of all deterministic algorithms solving it. Then, for any $x \in \mathcal{X}$ and $A \in \mathcal{A}$, the cost of executing $A$ on $x$ is denoted by $cost(a, x)$.

**Theorem E.1** (Yao's principle [34]). *Let $p$ be a probability distribution over $\mathcal{A}$, and let $A$ be an algorithm chosen at random according to $p$. Let $q$ be a probability distribution over $\mathcal{X}$, and let $X$ be an input chosen at random according to $q$. Then*
$$\max_{x \in \mathcal{X}} \mathbb{E}_p[\text{cost}(A, x)] \geq \min_{a \in \mathcal{A}} \mathbb{E}_q[\text{cost}(a, X)].$$

**Theorem E.2** (Chernoff bound [26, Theorem 4.4 and 4.5]). *Let $X = \sum_i X_i$ be the sum of 0/1 independent random variables. Then,*

1. *for any $\delta > 0$,*
$$\mathbb{P}\left[X \geq (1 + \delta)\mathbb{E}\left[X\right]\right] < \left(\frac{e^\delta}{(1 + \delta)^{1 + \delta)}}\right)^{\mathbb{E}[X]}.$$

2. for any $0 < \delta \leq 1$,
$$\mathbb{P}[X \geq (1+\delta)\mathbb{E}[X]] \leq e^{-\mathbb{E}[X]\delta^2/3}.$$

3. for $R \geq 6\mathbb{E}[X]$,
$$\mathbb{P}[X \geq R] \leq 2^{-R}.$$

4. for $0 < \delta < 1$,
$$\mathbb{P}[X \leq (1-\delta)\mathbb{E}[X]] \leq \left(\frac{e^{-\delta}}{(1-\delta)^{1-\delta)}}\right)^{\mathbb{E}[X]}.$$

5. for $0 < \delta < 1$,
$$\mathbb{P}[X \leq (1-\delta)\mathbb{E}[X]] \leq e^{-\mathbb{E}[X]\delta^2/2}.$$

**Theorem E.3** ([21, Equation 10]). *Let $Y = \sum_{i=1}^m Y_i$ be the sum of $m$ i.i.d. random variables with $\mathbb{P}[Y_i = 1] = p$ and $\mathbb{P}[Y_i = 0] = 1 - p$. We have for any $\alpha \in (0,1)$ that*
$$\mathbb{P}[Y \geq \alpha \cdot m] \leq \left(\left(\frac{p}{\alpha}\right)^\alpha \left(\frac{1-p}{1-\alpha}\right)^{1-\alpha}\right)^m.$$